\documentclass[a4paper,USenglish]{lipics-v2021}
\usepackage{craigstyle}

\title{Definability and Interpolation in Philosophy}

\author{Johan van Benthem}{University of Amsterdam, Stanford University and Tsinghua University}{j.vanbenthem@uva.nl}{https://staff.fnwi.uva.nl/j.vanbenthem/}{}

\begin{document}

\maketitle

\abstract{This paper offers a light tour of occurrences of Craig's interpolation theorem and Beth's definability theorem in philosophy. There is no coherent research program with a central agenda around this  theme, but extending the discussion in~\cite{Benthem2008},  we trace some lines and identify some issues in establishing contacts at this interface between the realms of logic and the philosophy of logic, science, language and information, epistemology, and metaphysics. This topic can also be viewed as a case study for more general aspects of the logic--philosophy interface.}

\tableofcontents

\section{The Broad Ideas}

Beth's Definability Theorem says that, in first-order theories, implicit definability of vocabulary is equivalent with explicit definability. More precisely, given a first-order theory T in a language with vocabulary $\mathbf{L}$ plus some predicate letter P, implicit definability says that, given two models $\mathsf{M}, \mathsf{M}'$ for T which are identical qua domain of objects and interpretation of the vocabulary of $\mathbf{L}$, the predicate interpretations P$^{\mathsf{M}}$ and P$^{\mathsf{M}'}$ must also be identical. Put more loosely, fixing the interpretation of $\mathbf{L}$ in a model for T automatically also fixes that of P. In contrast, explicit definability says that there exists a first-order formula $\varphi(\textbf{x})$ using only the vocabulary $\mathbf{L}$ with a tuple of variables \textbf{x} of the arity of the predicate P such that the theory T implies the formula $\forall \textbf{x} (P\textbf{x} \leftrightarrow \varphi(\textbf{x}))$. 

\vspace{-0.5ex}

\begin{theorem} For first-order theories, implicit definability and explicit definability are equivalent. \end{theorem}

\vspace{-0.5ex}

Clearly, explicit definability implies implicit definability, but Beth's achievement was to show the converse, which is not at all trivial for first-order logic~\cite{Beth1953}.

The background to Beth's Theorem was his interest in logic as a joint study of definability and proof, a tandem view of the history of the discipline since Antiquity on which he elaborated in various publications, in particular his magnum opus \cite{Beth1959} which also contains his more general philosophical views based on logical foundational research. In particular, Beth arrived at the Definability Theorem by reflecting on Padoa's Method for disproving explicit definability in the foundations of geometry. This combination of logic and philosophy of science was characteristic for the 1950s. Beth founded the ``Instituut voor Logica en Grondslagen van de Exacte Wetenschappen'' (Institute for Logic and Philosophy of the Exact Sciences), the precursor of the current Institute for Logic, Language and Computation in Amsterdam, and equally significantly, around the same time Beth's correspondent and ally Alfred Tarski founded the still existing Group in Logic and Methodology of the Exact Sciences at the University of California at Berkeley.
 
For our later discussion, it is convenient to rephrase the earlier formulation of Beth's notion to the following  equivalent version of implicit definability:

\vspace{-1ex}

\begin{quote} If two models $\mathsf{M}, \mathsf{N}$ both make T($\mathbf{L}$, P) true and F is an $\mathbf{L}$-isomorphism between $\mathsf{M}$ and $\mathsf{N}$, then F is also a P-isomorphism.\end{quote} 

\vspace{-1.2ex}

This isomorphism-lifting version  of Beth's Theorem generalizes to  weaker structure  invariances such as potential isomorphism\footnote{A \emph{potential isomorphism} between two models $\mathsf{M}, \mathsf{N}$ is a family {\bf F} of finite partial isomorphisms between these two models satisfying the following Back-and-Forth conditions. For every $f \in {\bf F}$ and $a \in \mathsf{M}$ there is a $b \in \mathsf{N}$ such that $f \cup \{(a, b)\} \in {\bf F}$, and the same condition holds in the opposite direction.}, a notion that will become important in Section 7 below. 


\vspace{1ex}

Ideas of implicit and explicit determination in this vein can be found in many areas of philosophy.

\medskip

Next let us consider the Craig Interpolation Theorem for first-order logic, \cite{Craig1957a,Craig1957b}, using the following simple formulation displaying just a few predicates for convenience. 

\begin{theorem} If formula $\varphi(P, Q)$ in predicate vocabulary \{P, Q\} semantically implies formula $\psi$ with vocabulary \{Q, R\}, then there exists a first-order formula $\alpha$ using only the shared vocabulary \{Q\} such that $\varphi(P, Q) \models \alpha(Q) \models \psi(Q, R)$. \end{theorem}

More elaborate versions of the Interpolation Theorem, of which there exists quite a few by now, cf. \cite{Otto2000} for a sample, give more detailed information about how the shared predicate Q occurs syntactically in $\alpha$ in terms of how it occurs in the formulas $\varphi$ and $\psi$. See   \refchapter{chapter:firstorder} in this book for more details and in-depth results.

{\it Note.} In various chapters of the present volume, a distinction is made between plain and deductive interpolation as versions of the preceding result which are not always equivalent for logical systems. While important, this distinction will not be highlighted in what follows. 

\vspace{1ex}

Craig Interpolation implies Beth Definability for first-order logic by a well-known simple argument. The converse does not hold. For instance, the  natural Guarded Fragment of the first-order language satisfies the latter property, but not the former, cf. \cite{HooglandMarxOtto1999,HooglandMarx2002}.\footnote{The first of these papers makes a case that the Guarded Fragment does satisfy the right version of interpolation from a `modal'  perspective -- an issue that we do not go into here.} For more information on these matters, see also  \refchapter{chapter:separation} in this book.

By an easy argument, Interpolation also implies the stronger Projective Beth Theorem which says the following, in a slightly simplified notation. If the theory T(A, B, C) $\cup$ T(A, B', C') implies $\forall$x (Cx $\leftrightarrow$ C'x), then T(A, B, C) implies an explicit definition for C by a first-order formula using the vocabulary A only. We will have a need for this stronger version occasionally.

\vspace{1ex}

We will also encounter applications of Interpolation in philosophy, though at first glance it may look like a more technical system property. For greater unity in philosophical applications of the two results discussed here, it would be nice if one could capture the surplus of interpolation over Beth in a natural and appealing manner. We will not resolve this matter here.\footnote{One candidate for a mathematical notion that might throw light on this comparative surplus is the  \emph{amalgamation property} in varieties of algebras, perhaps in a generalized category-theoretic setting.}

\vspace{1ex}

We are going to show how the above circle of ideas returns in philosophy in various forms. Section 2 is about the role of the two theorems in defining the very notion of a logical system, a live topic in the philosophy of logic. After that, the focus is on Beth's Theorem in Sections 3 -- 6 on supervenience, determinism, and reasoning about questions, culminating in the idea that the common thread in philosophical applications of Beth's Theorem  is the role of implicit and explicit \emph{dependence}. 

Then the main focus of our narrative shifts to Craig's Theorem and its role in thinking about the importance of \emph{vocabulary} in reasoning and theory construction, in Sections 7 -- 9 on generalized notions of logical inference, the architecture of scientific theories, and the modular structure of beliefs. We end with Section 10 flagging some connections with the  work of Bernard Bolzano, a creative and perhaps somewhat neglected 19th century pioneer in the emergence of modern logic.

\section{What is a Logical System?}

One immediate road into probing the importance of our two theorems is their relevance to a recurrent issue in the philosophy of logic, and just as well in logic itself. \emph{What is a logical system}, or in a more refined version: what is a good, or well-designed, logical system? This is not just an abstract quibble, what is at stake here is the proper perspective of what modern logic in its prolific system-building guise is about. What do our two theorems have to say here?

\vspace{1ex}

Our discussion will focus on the Interpolation Theorem, but the Definability Theorem plays a similar role. Satisfying a Craig-style interpolation theorem is sometimes quoted as a key desideratum in folklore discussions, since it expresses a form of `expressive completeness' of the logic for its language, as being able to formulate definitions and state crucial `transitions' from premises to conclusions. Even so, this intuition is not easy to make precise beyond just restating what the theorems say, and among formal approaches that go further, I only know of \cite{Feferman2008} and an earlier proof-theoretic analysis in \cite{Zucker1978}. A  general semantic way of seeing the centrality of Beth and Interpolation might go via their role in Abstract Model Theory, where Beth's Definability Theorem seems to follow from Lindstr\"{o}m's Theorem  for first-order logic under quite general conditions, though the case of Craig is less clear -- cf.  \refchapter{chapter:modeltheory} in this volume.\footnote{One might  question this criterion  given the ubiquity of failures of interpolation in important  systems found  in philosophical logic, cf.  \refchapter{chapter:nonclassical}  in this book. This issue will return  in Section 3 below.}

\medskip

But here is another way into the above question, restated as follows with a flavor of the crucial notion of invariance that we already noted in connection with Beth's Theorem:

\vspace{1.5ex}

\emph{When are two logical systems the same?} 

\vspace{1.5ex}

One answer,  found in both technical and philosophical literature, is this. The two logics $\mathfrak{L}_1$, $\mathfrak{L}_2$ should have mutual faithful syntactic \emph{translations} or embeddings, in the sense that the following equivalence holds for all formulas under some syntactic translation $\tau$ for the languages of these logics:

\vspace{-0.5ex} 

 \begin{quote} $\varphi \models \psi$ in $\mathfrak{L}_1$ \, iff \,$\tau(\varphi) \models \tau(\psi)$ in $\mathfrak{L}_2$
 
 \end{quote}
 
 \vspace{-0.5ex}
 
\noindent  and vice versa in the opposite direction from $\mathfrak{L}_2$ to $\mathfrak{L}_1$  for some syntactic  translation $\sigma$. 

\vspace{1ex}

This is just a start, however, and there is no philosophical consensus on what more might be required  for a good notion of system identity. Here is one possible richer view.

In Abstract Model Theory, logical systems come with both sentences and semantic structures, connected by a truth relation, and translation then also involves a semantic  transformation T of models running in the opposite direction, leading to the contravariant equivalence 

\vspace{-0.8ex} 

\begin{quote}  $\mathsf{M} \models \tau(\varphi)$ \, iff \, $\textrm{T}(\mathsf{M})\models \varphi$ \quad \quad (\#)\end{quote} 

\vspace{-0.5ex} 

This equivalence is in fact highlighted as the basic mechanism of information flow in Situation Theory~\cite{BarwiseSeligman1997} -- see Section 7 below for more on this framework. 

The abstract mathematical theory of this setting involves adjoint maps in Chu Spaces~\cite{Benthem2000}, an elegant category-theoretic perspective that we will not pursue here.\footnote{Faithfulness of the translation requires extra conditions, such as surjectivity of the transformation T.}

\vspace{1ex}

The broader issue at stake here is this. Looking under the surface of the myriads of logical systems  in the literature, what are the underlying invariances and connections?\footnote{One  famous   connection is the faithful \emph{G{\" o}del translation} from intuitionistic propositional logic into the classical modal logic $\mathsf{S4}$ which has led to a wide variety of fruitful contacts between the two systems.}  In other words, what is the coherence of the field of the logic? And again, do Craig and Beth have a bearing on this?

\vspace{1ex}

Without entering into a general philosophical discussion of these issues, here is one connection with our main topic in this paper. If Interpolation is made an integral part of the definition of an abstract logic, \emph{what constraint is induced on translations} between logical systems? Here, to  formulate the interpolation property in its proper generality, we need abstract notions of `sentences', `consequence', `vocabulary', and `occurrence'. Proper translations should now respect this basic structure, and make sure that  interpolation  is transferred from one logical system to any properly translation-related logical system, a concrete constraint which now becomes a more precise issue.\footnote{As a concrete illustration of this transfer, \cite{BenthemCateKoudijs2022} show how a pair consisting of suitable homomorphic translation from the Guarded Fragment  into the Modal Dependence Logic LFD mentioned in Section 9 below, plus a converse satisfiability reduction yields the Interpolation Theorem for LFD from the earlier-mentioned Beth Theorem for the Guarded Fragment. More such case studies are needed to see more clearly which constraints are imposed by transfer of Interpolation.}

\vspace{1ex}

Still, making  Interpolation  key to the identity of logical systems and translations between them does not  establish just \emph{why} one should consider this  property so important. \cite{BenthemDoets1983} conjectured that Interpolation is the key axiom in a \emph{complete axiomatization of the meta-theory} of first-order logic, viewed as a two-sorted structure consisting of sentences and finite vocabulary sets with relations of logical consequence and `occurrence'. The thinking behind this conjecture was that Craig interpolation seems the last major property of first-order logic discovered in modern history.\footnote{Of course, this historical observation is no longer true when we consider the remarkable probabilistic properties of first-order logic discovered later, such as the Zero-One Law, \cite{EbbinghausFlum1995}.}  However, \cite{Mason1985} showed that even the meta-theory of classical propositional logic construed in this way is \emph{non-axiomatizable}, and in fact admits of a faithful embedding of True Arithmetic. Thus, the central role of Interpolation does not show in this particular construal of the structure of logical systems -- but perhaps all this shows is that we have made the notion of `meta-theory' too concrete, importing peculiarities of the natural numbers and the complexity of their theory.

\vspace{1ex}

We conclude this section with another link to our main theme that may be of interest in itself. The topic of translations also  raises an interesting connection with the Beth Definability Theorem. 

\vspace{1ex}

\textbf{Translation as generalized definability} \, One can think of an explicit \emph{syntactic translation} $\tau$ satisfying the above-mentioned  equivalence ($\#$) as providing a \emph{generalized explicit definition} for the basic notions of the logic $\mathfrak{L}_1$, not inside of its own vocabulary (the special case of definability at issue in Beth's Theorem) but in terms of formulas in another language:  namely, that of the logic $\mathfrak{L}_2$. Accordingly, this time, the corresponding notion of \emph{implicit definability} is not in just one model of the same similarity type, but it runs across possibly different models. 

Here is a first intuitive formulation of this idea:

\vspace{1ex}

\quad  Fixing the $\mathfrak{L}_2$-structure of a model $\mathsf{M}$ suffices for fixing the denotation of  

\quad the basic predicates of the logic $\mathfrak{L}_1$ in all T-matching models $\textrm{T}(\mathsf{M})$ for  

\quad transformations T satisfying some suitable conditions. 

\vspace{1ex}

More precisely, let the transformation T in the above  definability schema ($\#$) be defined  as a surjective  function from $\mathfrak{L}_2$-models to $\mathfrak{L}_1$-models which also maps the domain of a $\mathfrak{L}_2$-model onto the domain of the matching $\mathfrak{L}_1$-model, i.e., T also maps objects to objects. \emph{Note:} this is surjectivity for objects in domains of T-correlated models, not for T as a map from models to models themselves. 

It is easy to see that the schema ($\#$) then implies the following cross-model condition, which we state here for unary predicates only, purely for convenience:

\vspace{1.5ex}

\quad	If T: $\mathsf{M} \rightarrow \mathsf{N}$ with T(b) = a, $\mathsf{M}$ is $\mathfrak{L}_2$-isomorphic to $\mathsf{M}'$ with b matching b', and \vspace{0.5mm}

\quad T: $\mathsf{M}' \rightarrow \mathsf{N}'$ with T(b') = a', then for any $\mathfrak{L}_1$-predicate P, P$^{\mathsf{M}}$(a) implies P$^{\mathsf{M}'}$(a').

\vspace{1.5ex}

\emph{Technical elaboration.} This setting invites a use of the Projective Beth Theorem, or rather, the Interpolation   Theorem behind it. It applies to yield an explicit translation from an implicit one, provided the above transformation T is defined in first-order terms. Let this definition take the form of a first-order theory $\Sigma$  with vocabulary \{$\mathbf{L}_1, \mathbf{L}_{2}$, T, M, N\} where  M, N stand for disjoint unary predicates, the $\mathbf{L_1}$-predicates (the vocabulary of the logic $\mathfrak{L_1}$) can only hold inside the subset defined by N, the $\mathbf{L}_{2}$-predicates (from the logic $\mathfrak{L}_{2}$) can only hold inside the subset defined by M, and where $\Sigma$ only makes assertions about objects in M $\cup$ N, even if a model itself is larger.\footnote{This effect can be achieved by assuming that all quantifiers in $\Sigma$ are  bounded by M $\lor$ N. The  `book-keeping conditions' stated here do not seem overly restrictive for defining a function between models.}

Now we formulate the above notion of extended implicit definability as a first-order consequence in an extended language, where P is one predicate from $\mathbf{L}_{1}$ (taken to be unary again, just for simplicity), a, a', b, b' are individual constants, and Z stands for a binary relation:

\vspace{1.5ex}

$\Sigma(\mathbf{L}_{1}, \mathbf{L}_{2}$, T, M, N), Tb = a, Pa (i), and \vspace{0.5mm}

``Z is an $\mathbf{L}_{2}$-isomorphism between M and M', b Z b'' (ii), and \vspace{0.5mm}

$\Sigma(\mathbf{L}_{1}$', $\mathbf{L}_{2}$', T', M', N'), T'b' = a' (iii) \vspace{0.5mm}

$\models$ P'a'.

\vspace{1.5ex}
This says that if we have two pairs of transformation-related models, where one source model is $\mathbf{L}_{2}$-isomorphic to the other, the target models agree on the interpretation of the $\mathbf{L}_{1}$-predicate P. 

\vspace{1ex}
 
By a standard appeal to  the Compactness Theorem for first-order logic, the total set of formulas occurring in the clauses (i), (ii) and (iii) here can be taken to be finite, and with some abuse of notation we can rewrite this consequence to one that holds between formulas:

\vspace{1.5ex}

$\Sigma(\mathbf{L}_1, \mathbf{L}_{2}$, T, M, N) $\land$ Tb = a $\land$ Pa $\models$ (``Z is an $\mathbf{L}_{2}$-isomorphism between M and M''  \vspace{1mm}

 $\land$ b Z b' $\land$ $\Sigma(\mathbf{L}_1$', $\mathbf{L}_2$', T', M', N')  $\land$ T'b' = a') $\rightarrow$ P'a'.

\vspace{1.5ex}
 
Here the only overlap in vocabulary  between antecedent and consequent  is the set \{M, $\mathbf{L}_{2}$, b\}. Therefore, there is an interpolant $\delta$(M, $\mathbf{L}_{2}$, b) in this vocabulary which just describes a property of the object b in M-type models.\footnote{Reducing to this syntactic form may take some first-order manipulation, noting that quantification over N in the formula $\delta$ has no essential effect since no L$_2$-predicates can be true there.} 

\vspace{0.5ex}

By some standard manipulation, we then see that the following two assertions hold:

\vspace{1.3ex}

(a) \, $\Sigma(\mathbf{L}_1, \mathbf{L}_{2}$, T, M, N) $ \models$ Tb = a $\rightarrow$ (Pa $\rightarrow \delta$) \vspace{1mm}

(b) \, (``Z is an $\mathbf{L}_2$-isomorphism between M and M'' $\land$ b Z b' $\land$ $\Sigma(\mathbf{L}_{1}$', $\mathbf{L}_{2}$', T, M', N')  $\models$  \vspace{0.5mm}

\quad\quad  Tb' = a'  $\rightarrow (\delta \rightarrow$ P'a')

\vspace{1.2ex}

 The identity relation on M satisfies all conditions for an $\mathbf{L}_{2}$-isomorphism between M and M' = M. Substituting this for the predicate Z in clause (b), while renaming some predicate variables, implies

\vspace{1.5ex}

\, $ \Sigma(\mathbf{L}_{1}, \mathbf{L}_{2}$', T, M, N)  $\models$  Tb = a  $\rightarrow (\delta \rightarrow$ Pa).

\vspace{1.5ex}

 Thus, in all, the theory $\Sigma$ implies the explicit definition   \vspace{1.5ex}
 
 \quad\quad Ta=b $\rightarrow$ (Pa $\leftrightarrow \delta$(a, b)).
 
  \vspace{1.5ex}
  
 Finally, given $\mathbf{L}_{2}$-definitions for the  predicates in $\mathbf{L}_1$, a standard inductive proof lifts this to an explicit \emph{compositional translation} for all formulas of $\mathfrak{L}_1$. \emph{Note:} Here the  quantifier step uses  the above assumption that T is a  surjective function from the domain of M onto that of N.\footnote{Another construal of  implicit definability would fix the map T in the assertions about M, N and   M', N'.  Interpolation still applies, but we  now also get undesirable definitions for the predicate P that do not just refer to $\mathbf{L}_2$, but also to the function T, as in ``Pa holds iff exactly 2 objects in M get T-mapped onto a''.}

 \vspace{1ex}
  Of course, the preceding construal does not yet analyze the notion of `translatability' as a  relation between logical systems, which does not start from a given semantic transformation. For an abstract analysis of the latter sort, cf.~\cite{BenthemPearce84}.

\vspace{1ex}

In summary, this section has established some connections between our key theorems and issues in the philosophy of logic that may be worth reflecting on, but no definite answers were provided that could not be challenged. This sort of delicate contact is typical for the logic-philosophy interface, and we will encounter further instances as we proceed.

\vspace{1ex}

Our discussion in this section has focused on isomorphism as the central semantic invariance. This is just one choice, and for instance, in Section 7, we will also use the weaker notion of `potential isomorphism' introduced earlier. But in principle, our  discussion would also apply to a whole spectrum of invariance notions, including still weaker ones such as \emph{bisimulation}, in which case the results we cited would be talking about definability in \emph{modal}  rather than complete first-order languages.

\section{Supervenience}
Our next topic starts from semantic invariance relations and their liftings. In philosophy, the notion of \emph{supervenience} of a predicate Q on one or more predicates P is ubiquitous.  Examples with broad sweep are the supervenience of our mental states on our physical/biological states, of the moral aspects on the factual aspects of a scenario, and so on. 

For an excellent survey, we refer to the Encyclopedia article ~\cite{McLaughlin2023}, from which we take just a  few notions to make our main points. For a start, let us say that Q \emph{supervenes on} P if

\vspace{-1.5ex}

\begin{quote} \emph{Whenever two situations are} P-\emph{equivalent, they are also} Q-\emph{equivalent}.\end{quote} 

\vspace{-1.4ex}

We state this for situations, but we can also think of supervenience as comparing objects, or whatever relevant entity we want to consider. One reason why supervenience is attractive is that it expresses a `harmony' qua properties between two different realms, but not necessarily brute-force  \emph{reducibility} of Q to P via some explicit definition `explaining Q away'.

Of course, formulations like this require further precision: which realm of situations or objects is considered, and which precise notion of semantic equivalence is meant? But even without specifying these parameters further, it should  be obvious that  supervenience has the same flavor as implicit definability in the sense of Beth's Theorem.\footnote{As we shall see in Section 9, supervenience is a form of  abstract \emph{dependence}, and in fact some authors have analyzed its general logic in these terms, cf. \cite{Fan2019}.}

In fact, more can be said, since several  varieties of supervenience, weaker and stronger, have been proposed in the philosophical literature. Here are just two samples. In a \emph{local version}, the formulation that we started with   compares equivalent or similar items inside one given world. Transposing this to a model-theoretic formulation, one way of expressing this more formally is that in a given model, P-automorphisms are Q-automorphisms – though other similarity relations can also be used. 

But in a more \emph{global version} across different worlds, the requirement might be as   follows: 

\vspace{-1.5ex}

\begin{quote} \emph{If object} x \emph{in world} w \emph{is} P-\emph{similar to object} y \emph{in world} v, \emph{then} x, y \emph{are also} Q-\emph{similar}.\end{quote}

\vspace{-1.5ex}

\noindent In model-theoretic terms, this has precisely the flavor of the notion of implicit definability as introduced in Section 1, with similarity  cashed out as isomorphism or potential isomorphism.

\vspace{0.8ex}

One recurrent  debate in this area of philosophy has been this: When does supervenience imply \emph{reduction} in the sense of explicit definability? This is indeed what Beth's Theorem would say, but to apply this result to supervenience, one must first introduce a formal first-order language plus a background theory into the philosophical discussion. It might be debatable whether this methodological constraint  fits the spirit of the philosophical literature, or at least, the felicity of this formalization move needs argumentation. As we said, 
failure of explicit definition is in fact one reason why supervenience is philosophically attractive as a non-reductionist position positing merely harmony between two realms without brute force reduction. 

Two points can be made here. First, some logics with the Beth property might still be harmless from a reductionist perspective. For instance, in second-order logic an explicit definition can be extracted routinely from an implicit one by means of an impredicative existential quantification over the irrelevant vocabulary, trivializing Beth's Theorem. Such a `reduction' seems harmless. 

But more interestingly, one might try to take a fresh look at the relevant philosophical literature using non-higher-order non-classical logics that \emph{lack the Beth property} as a way of avoiding reductionism. One example are the relevant logics lacking the Beth property discussed in \cite{Urquhart2025}, cf. also  \refchapter{chapter:nonclassical}  in this book.  But of course, this weakening of the base logic should also have some independent motivation: mere avoidance of the Beth property is hardly a positive virtue. After all, the function is there, but the vocabulary in logics of this kind does not allow us to define it explicitly. One immediate test for this way-out would be if the known technical counterexamples (sometimes produced by computer-assisted search) have any independent conceptual interest.\footnote{Even so, somewhat perversely, could it be that undefinability (and once on this track, unprovability), i.e., keeping some things implicit only, might sometimes be regarded as positive features of a logical system?}

\vspace{0.8ex}

But there is also an alternative way of thinking that calls for different distinctions. By the above definition, supervenience as implicit semantic definability already says that there is an \emph{abstract functional dependence} of the supervening predicate Q as the value of some function  on the equivalence classes for its underlying predicate P.  Beth's theorem then adds that, under special circumstances, this abstract functional dependence can be expressed generically across models, by some explicit first-order formula of the P-language. What does this further feature really imply?

Returning to  the  philosophical concern of brute-force `reducibility',  even such an explicit definition need not be a burdensome form of reduction if that definition is so unwieldy as to be inaccessible to bounded human agents. In this sense, one might say that explicit reduction only hurts and constrains us if we (or others) can \emph{grasp and manipulate} the dependence. 

What such finer distinctions would call for are more computational analyses of the powers of bounded human agents, and in terms of Beth's Theorem, stricter bounds on the \emph{complexity of the definitions} produced. In this way, some of the work in computational logic on Beth-type theorems presented in  \refchapter{chapter:proofcomplexity} and \refchapter{chapter:automated}  might become relevant to philosophical discussions.

\vspace{1ex}

Finally, interestingly, the earlier local version of supervenience is not directly the content of Beth's Definability Theorem, and rather falls under this well-known theorem  from \cite{Svenonius1959}: 

\vspace{1ex}

If $\mathbf{L}$-automorphisms are also P-automorphisms in all models for a first-order theory T, then T implies a disjunction of explicit definitions for P in terms of $\mathbf{L}$, not necessarily one single definition.\footnote{To see the difference, think of a theory T consisting of just the statement $\forall x(Px \leftrightarrow Lx) \lor \forall x (Px \leftrightarrow \neg Px)$. This does not satisfy the implicit definability for Beth's Theorem. Two models for T with the same domain and denotation of the predicate $L$ can still disagree on the denotation of $P$. But what is true is that in each of these  models, the automorphisms  that leave $L$ in place will also leave $P$ in place.} 

\vspace{1.2ex}

Supervenience is just one instance where a more detailed study of a rich philosophical literature might match up with a greater variety of notions and results in existing logical theory. We now turn to other examples of such interfaces.

\section{Determinism}
The notion of supervenience also relates to another broad philosophical theme: that of \emph{determinism}. This notion occurs in classical metaphysics, where  Aristotle's famous Sea Battle argument investigated a sense in which the logical law of Excluded Middle for future truths seems to imply determinedness of our present behavior placing it outside of our control. Similar themes have kept returning, all the way to causal determinism in the empirical sciences, though the determination there usually runs in the opposite forward temporal direction, from what is true now to what will become true later. 

A lot is at stake in debates around this pervasive notion: ranging from the (non-)existence of free will to the role of causality in Artificial Intelligence or the nature of what physics can provide.

\vspace{1ex}

As with supervenience, many proposals have been made for making the notion of determinism more precise including analyses by prominent philosophers such as Bertrand Russell. 

\vspace{1ex}

As it happens, in this area, there is a landmark attempt by a well-known logician to sort out the issues and apply Beth's and Craig's results in the philosophy of science. \cite{Montague1974} formalizes physical theories such as Newtonian mechanics in first-order terms as having a mathematical part referring to the  natural numbers $\mathbb{N}$ and real numbers $\mathbb{R}$ plus tuples of functions on reals that encode  relevant physical system properties such as position and momentum of particles, and other  notions -- including abstract, non-observable ones postulated by the  theory.

In Montague's analysis, a physical theory T describes a set of possible histories, where we think of models as having the standard natural and real numbers (here the latter also encode the passage of time), while functions defined on the reals give the current state of, say, a mechanical system as a function of the time. Then different definitions of can be given of what is a `deterministic theory', reflecting various proposals in the philosophical literature which we will not reproduce here.\footnote{Montague notes that theories can easily be deterministic in terms of arbitrary functions for the dependence of one state on others, while the more practical issue is generic definability of the functional dependence, fit for  computational uses. This distinction between model-dependent definitions and generic ones for all models echoes an earlier point made about supervenience and reductionism in Section 3.} 

\vspace{1ex}

Here is one typical `implicit' version which  occurs widely in the literature, going back to a well-known historical statement of determinism in physics formulated by Laplace:

\vspace{-1ex}
\begin{quote} \emph{If two histories allowed by} T \emph{have equal states for the physical system at time} t$_0$, \emph{and} t$_0$ < t, \emph{then the two histories also have the same total system state at time} t. \end{quote}

\vspace{-1ex}

Modulo quite a few syntactic details in Montague's treatment which are irrelevant to the intended illustration here, this semantic condition on models for physical systems can be formulated as a form of implicit definability in a first-order theory. Craig Interpolation and Beth's Theorem then apply to conclude that the theory also provides strong determinism in the form of an explicit definition in the language of the theory for the system state at time t in terms of the state-at-time-t$_0$, t$_0$ and t which can be used for physical reasoning and calculation.

At the same time, linking to our earlier discussion of various perspectives on supervenience, Montague's formal apparatus also allows us to see more precisely what is required for the Beth property, and thus, indirectly also, how one might try to avoid it.

\vspace{1ex}

It has to be said that there is a lot  of  fine-print to this analysis, and Montague's paper may not have had the impact it deserves due to  an overload of notions, results and conclusions pointing in various directions. However, this can again be seen as an illustration of an earlier general point: bringing logic to bear on philosophical notions need not lead to unique or definitive outcomes. Also, our earlier point applies that the logical results can only be made to work if one first formalizes the general philosophical discussion of determinism in terms of formal languages and theories.

\vspace{1ex}

There are many further types of formalization of empirical theories by logical means, on which there exists a vast literature, cf. \cite{Suppe1974,Benthem1982}. Of the many classical themes treated in this literature, one will return in Section 8 below, viz. the different vocabularies that go into the making of an empirical theory, where  the relevant connection is now with Interpolation. 

\vspace{1ex}

For now, we move to another broad topic in philosophy in the spirit of our themes so far.

\section{Questions}

New information, whether obtained by experiments in scientific theories or through more mundane channels of communication, often arises from asking \emph{questions}. Accordingly, questions are ubiquitous in the philosophy  of  language, or in general epistemology -- where knowledge can be seen as knowing the answer to some relevant question which controls the direction of inquiry. 

\vspace{1ex}

Ever since the early pioneering work documented in~\cite{Ajdukiewicz1978} it has been proposed by philosophers to let logic deal with, not just consequence relations among propositions or statements that give information, but also between questions that ask for information. 

In a very common setting for implementing this, cf. \cite{GroenendijkStokhof1997}, logical semantics for questions models them in terms of their complete answers as forming a partition of a relevant state space. One often thinks of the partition as a current `issue' to be resolved by identifying one of the partition cells as the actual one.

Valid consequence from a set of questions then becomes the following notion, which has been around for a long time in the semantic literature: 

\vspace{-1.3ex}

\begin{quote} \emph{If we have all the answers to the `premise' questions} ?P$_1$, ..., ?P$_k$, \emph{we automatically} 

\emph{also have an answer to the `conclusion' question} ?C. \end{quote}

\vspace{-1ex}

For instance, in this sense, (i) ?A implies ?$\neg$A, (ii) ?A, ?B imply ?(A \& B), and so on, suggesting a connection between  consequence for questions and definability from premises. 

\vspace{1ex}

A semantic definition for the notion underlying consequence between questions is as follows on modal state spaces where formulas can be true or false: 

\vspace{-1ex}

\begin{quote} \emph{If two states} s, t \emph{agree on all}  P$_i$-\emph{truth values} (1 $\leq$ i $\leq$ k), \emph{they also agree on the} C-\emph{truth value}. \end{quote} 

\vspace{-1ex}

\noindent Stated in other terms, the joint refinement of the P$_i$-partitions is included in the C-partition.

\vspace{1ex}

 Of course, there is much more to the actual behavior of questions in natural language, as they raise or modify issues and steer the course of conversation and inquiry. In addition, questions interact with other informational events, and for instance,  the above notion of question-consequence can be made conditional on events !Q of receiving the indubitable information that Q by restricting the total state space only to states where Q holds. 
 
\vspace{0.8ex}

The preceding notion is clearly in the same spirit as the invariance and similarity notions discussed earlier in connection with Beth-style implicit definability.\footnote{In fact, analogies run much wider, and similar `equal answer' equivalence relations have been used to study the widespread phenomenon of `ceteris paribus' reasoning~\cite{BenthemGirardRoy2009}.} Using this analogy, a complete logic for consequence between questions suitably defined in a predicate-logical setting was given in \cite{CateShan2007}, who use the slightly strengthened projective version of Beth's Theorem to prove the following key equivalence for conditional implication between questions: 

\vspace{-0.5ex}

\begin{theorem} !R, ?P $\models$ ?Q \, iff \, there exists a first-order formula $\alpha$(P) such that R $\models$ Q $ \leftrightarrow \alpha$(P). \end{theorem}

\vspace{-0.5ex}

This result can be seen as a sort of expressive completeness of the usual logical constants for the purpose of giving explicit answers, typing logic of questions closely to logic of propositions.

Also, interestingly, the cited paper   connects Uniform Interpolation (see \refchapter{chapter:uniform} in this book) with the existence of a `best answer' to a question given some background information.

\medskip

Thus, the philosophically important notion of questions can be brought within the scope of logic, and the Beth and Craig theorems are key to obtaining more definite results.\footnote{For much more on the still growing area of logics of questions, we refer to~\cite{CrossRoelofsen2022}.} 

\medskip

Again, this is just a start, and many issues remain for discussion, and here is just one of them. The preceding account will work in a first-order setting, but it is of course a moot point whether natural language is first-order. In particular, do the above results go through for extended languages with non-first-order \emph{generalized quantifiers} such as ``most'' or ``many''? 
In this area of Generalized Quantifier Theory, as elsewhere, formal semantics of natural language meets with issues in mathematical logic concerning the properties of logical systems, \cite{PetersWesterstahl2006}.

\section{Dependence}

It is time to step up the abstraction level of our analysis. Consequence between questions is a special case of semantic \emph{functional dependence} between variables whose values can be truth values for Yes/No-questions, or other objects in the case of more general \emph{wh}-questions, cf. \cite{Ciardelli2016}. But functional dependence also lies behind many notions discussed earlier in this paper, with our analysis of determinism involving variables taking values in state space, or supervenience as a case where the values of variables are even just equivalence classes of some relevant invariance relation.

\vspace{1ex}

In modern dependence logics, \cite{Vaananen2009}, functional dependence runs between sets of variables X and variables y in \emph{state spaces} $\mathsf{M}$ where not all simultaneous assignments of values to these variables may be available at states. These `gaps' are precisely what induce correlations between the values of variables, of which functional dependence is one important species.\footnote{Note that, in this semantic setting, variables are no longer the anonymous binding devices used in standard first-order logic, but named functions from states to values, displaying an individual behavior reflecting the uses of the notion of `variable' in empirical science or probability theory.}

\vspace{1ex}

The  dependence setting also makes sense in other areas of philosophy, such as the Situation Theory to be discussed in Section 7, where information flows and knowledge arises because of various types of informational correlation between situations in the world, which  need not come from their structural similarities, but  be enforced by global constraints on the behavior of the overall system.

\medskip

The above standard semantic definition of functional dependence is one of \emph{implicit semantic determination}, where we say that s =$_X$ t if s, t give the same values to all variables in the set X. Here is the general  notion of functional dependence of a variable $y$ on a set of variables $X$:

\vspace{-1.2ex}

\begin{quote} \emph{For any two states} s, t \emph{in state space model} $\mathsf{M}$, s $=_X$ t \emph{implies} s =$_y$ t. \end{quote}

\vspace{-0.8ex}

This abstract framework applies to questions with truth values for the relevant propositions, but also to knowing objects  as in constructions like ``knowing who killed Mr. T", and even to the earlier notion of supervenience. It also applies to implicit definability in Beth's sense when values of variables can be whole equivalence classes of some relevant similarity relation.

The earlier-quoted completeness theorem for the basic logic of propositional questions can then be seen as special case of a general minimal propositional modal logic of functional dependence proposed in~\cite{BaltagBenthem2021} with questions as variables and truth values as their values.

\medskip

In this setting, an earlier point returns. The stated semantic equivalence obviously suffices for the existence of an abstract function  F on value domains for variables so that for each state s, 

\vspace{-1ex}

\begin{quote} s(y) = F(s[X]). \end{quote} 

\vspace{-1ex}

While such a function can always be found ad-hoc, locally in a given state space model $\mathsf{M}$, Beth's Theorem then gives a special definable  function F, under the right conditions, which works generically across models. This suggests a more general perspective on Beth-type results as an interesting companion to dependence logics, as providing conditions under which general abstract dependencies can be made explicit and thus fit for computation and deduction.

\medskip

Given these connections, it is a natural question whether current dependence logics satisfy Beth Definability or Craig Interpolation theorems with respect to variables, and not just shared predicate vocabulary. For  second-order dependence logic in the style of \cite{Vaananen2009}, the answer is that even Uniform Interpolation (see  \refchapter{chapter:uniform} for a definition and results) holds in the variable-sharing sense, as is explained in  \refchapter{chapter:modeltheory} in this book. For the decidable modal dependence logic of~\cite{BaltagBenthem2021}, only the predicate version of interpolation is known, \cite{BenthemCateKoudijs2022}, but it seems likely that a shared-variable version holds as well.

\vspace{0.5ex}

Dependence seems to be the common thread behind most of our philosophical topics in this paper. It clearly  applies to our discussions of supervenience and determinism. But dependence logic also applies to our earlier discussion of translations between logical systems, since as we noted, the existence of an explicit syntactic translation matches an implicit semantic dependence in the model classes for the two logics being connected. The reason why this works is, of course, that the variables in dependence logics can stand for anything, and likewise, their values can be anything from numbers or objects to equivalence classes of models or even information states of agents.
\vspace{0.5ex}

Given this ubiquity, basic dependence reasoning  is an interesting candidate for extending the core calculus of logic beyond the usual repertoire of `logical constants'  in standard logics. 

\vspace{0.5ex}

Going further, functional dependence of y on X is one end on a spectrum, where fixing the values of the variables in X leaves only one value for y. At the opposite end lies \emph{independence}, where  the current X-values place no constraint on the values that y can take at the current state. Independence, too, might play a  role as a  philosophical notion, say, as a sort of `anti-definability' or `unlearnability'. And in between on the spectrum lie notions of \emph{correlation} between X and y which can `soften' the force of functional dependence. This might make sense in various areas considered in this paper. 

This correlation perspective again suggests interesting mathematical issues as to whether Beth's Theorem and Craig's Theorem admit of `softer' versions producing definitions or interpolants up to exceptions, in either qualitative or even probabilistic senses.

\vspace{1ex}

We now leave the invariance/dependence thread, and look at how definability and interpolation highlight the crucial role of \emph{vocabulary} in shaping both everyday beliefs and scientific knowledge.

\section{Information and Generalized Consequence}
Let us start with a core notion studied in logic, viz. \emph{consequence}. A common philosophical view holds that drawing logical consequences has to do with \emph{information}, in particular, with elucidating information that is implicitly present in given premises or data. The handbook chapter~\cite{BenthemMartinez2008} discusses this view of consequence, while also identifying other broad notions of information at work in logic.

A broader focus on information is found in Situation Theory, \cite{BarwisePerry1983}, a formal paradigm inspired by the  view of knowledge as based on information flow in~~\cite{Dretske1983}. In this philosophical system the world consists of partial situations (at least if one takes the theory as one of metaphysics rather than epistemology), and information flows because of the existence of \emph{correlations} between the behavior of these situations, mediated by abstract or concrete physical channels. 

\vspace{0.5ex}

Logical consequence may now be thought of accordingly as acting  typically  \emph{across different situations} or, more formally, different models. Say, when looking at a radar screen on the ground, we observe light dots changing position, and conclude to movement of planes far away through a correlation between their actual flight paths and what is visible on the screen we are looking at.\footnote{This cross-situational function of inference was already present in ancient Indian logic with its standard example of `smoke means fire', drawing conclusions about fire on an inaccessible mountain top from observations made here on the ground, \cite{Bochenski1961}.}  

What the relevant connections are between the two situations in such inferences can of course vary in practice, since many kinds of correlation can carry information.

\vspace{1ex}

This situational view suggests a new notion of entailment, not just in one model as with ordinary logical consequence, but across models. Here are some notions and results from~\cite{BarwiseBenthem1999} developing this theme. We say that $\varphi$ \emph{entails} $\psi$ \emph{along a relation} R if, 

\vspace{-1ex}

\begin{quote} for any two models $\mathsf{M}, \mathsf{N}$, if  $\mathsf{M} \models \varphi$ and $\mathsf{M}$ R  $\mathsf{N}$, then   $\mathsf{N} \models \psi$.\end{quote} 

\vspace{-1ex}

\noindent On this perspective, a model $\mathsf{M}$ does not just support 
truths about itself which can be elucidated by inference from what we already know about $\mathsf{M}$, the model is also a beacon `broadcasting' further information to other models along various relations. These relations range from very strong, like isomorphism, to weaker, including homomorphism or modal bisimulation. The further away one gets from $\mathsf{M}$ qua structural similarity, the weaker the information transfer.

\medskip

This setting offers a fresh look at the Interpolation Theorem. We already have a valid consequence $\varphi \models \psi$. What is the further benefit of having an interpolant $\alpha$ in the shared vocabulary $\mathbf{L}$? What this really gives us is the following form of cross-model transfer: 

\vspace{-1ex}

\begin{quote} if  $\mathsf{M} \models \varphi$ and $\mathsf{N}$ is $\mathbf{L}$-potentially isomorphic with $\mathsf{M}$, then  $\mathsf{N} \models \psi$.\footnote{There are also other interpretations of the surplus. Say, one could view interpolation as stating that a logical consequence can always be related to a \emph{relevant} implication in the sense of shared topics.}  
\end{quote} 

\vspace{-1ex}

\noindent That is, thanks to the interpolant, $\varphi$ entails $\psi$ along $\mathbf{L}$-potential isomorphism.\footnote{Formally speaking, this notion has a modal form $\varphi \rightarrow [p.i._\mathbf{L}]\psi$. This also suggests looking at other modal notions such as $\langle p.i._\mathbf{L}\rangle \varphi$ which defines all pointed models that are $\mathbf{L}$-potentially-isomorphic to the current model. The latter would define a uniform interpolant in the present sense for the formula $\varphi$.} 

This analysis of what interpolants do inspires a new version of the Interpolation Theorem. 

\vspace{-0.8ex}

\begin{theorem} The following are equivalent for formulas as described just now: 

(i) Existence of an $\mathbf{L}$-interpolant, (ii) Entailment along potential $\mathbf{L}$-isomorphism. \end{theorem}

\vspace{-0.5ex}

For first-order logic, this new version is equivalent to the usual Craig Interpolation Theorem, using the downward  Löwenheim--Skolem theorem plus the fact that, on countable models, potential isomorphism and isomorphism amount to the same thing. But in general, the new formulation is weaker, and for instance, for the fundamental infinitary logic  $\mathfrak{L}_{\infty \omega}$ (see \refchapter{chapter:modeltheory}), often touted as a typical example of failure of the standard interpolation theorem, this new version turns out to hold.\footnote{The proof of this result uses a Lindstr\"{o}m-type argument with instead of the usual reliance on the Compactness Theorem, the  still very powerful model existence property encoded by the well-known \emph{projective undefinability of the notion of well-ordering} in the infinitary logic $\mathfrak{L}_{\infty \omega}$.} 

\medskip

This brings us back to a general issue raised in Section 2. When saying that a good logical system might have Interpolation as a characteristic feature, we took the standard formulation of Craig's Theorem for granted. But with our two options on the table, this  becomes more delicate. 

What is the natural formulation of the Interpolation property? And more generally, what are natural stable versions of known meta-properties of logical systems that we would like to transfer, discounting peculiarities of their original formulation when first discovered?

\vspace{1ex}

Similar generalized-interpolation versions exist for the classical model-theoretic \emph{preservation theorems} tying important types of semantic behavior to syntactic form, whose content already involves crossing between models. For instance, instead of the usual Los--Tarski theorem characterizing preservation under submodels,  we also have the following version:

\vspace{-0.5ex}

\begin{theorem} $\varphi$ entails $\psi$ along $\mathbf{L}$-submodels \, iff \, there exists a syntactically universal $\mathbf{L}$-formula $\alpha$

 which interpolates as follows:  $\varphi \models \alpha \models \psi$.\end{theorem}

\vspace{-0.6ex}

The original preservation theorem is retrieved by taking the special case that $\varphi$ = $\psi$. A similar case is the following characterization theorem for modal logic: 

\vspace{1.5ex}

A first-order  $\varphi$ entails $\psi$ along $\mathbf{L}$-bisimulation iff there is a modally $\mathbf{L}$-definable interpolant. 

\vspace{1.5ex}

The proofs of such results often proceed by  inspecting that standard proofs of the preservation actually also establish the interpolation version.

\medskip
There is more to the logical correlation view of information flow between situations, since it need not arise from structural similarity, but can  also result from global dynamical system behavior. This  phenomenon was already identified in our discussion of dependence logics in Section 6 above. 

\medskip 

What is the philosophical import of the preceding discussion? In our view, inspired by early work on Situation Theory, this section has addressed some fundamental issues in the philosophy of logic but also in the  general philosophy of information and one might say, even in epistemology.

\medskip

We now continue with other aspects of the role of vocabulary in philosophy, and in particular, the structuring of theories consisting of organized data and regularities that we reason with.

\section{Theoretical Terms in Scientific Theories}
One key theme in  the philosophy of science has been the difference between (a priori) mathematical theories and  theories in the empirical sciences. The latter contain \emph{observable terms} whose values can be observed or measured directly, but equally crucial are  \emph{theoretical terms} whose values cannot be observed directly, but which organize and streamline the theory. Where one draws this boundary in any given theory has been a matter of much philosophical debate, but the observational/theoretical distinction makes obvious intuitive sense, not just in scientific settings.

\vspace{1ex}

Worth noting historically here is the seminal paper \cite{Beth1949} which for the first time introduced the idea that empirical theories in science could be studied model-theoretically as combining syntactic formulations of laws with defining collections of structures representing actual physical systems. 

\vspace{0.5ex}

In these logical terms, consider a syntactic theory T($\mathbf{L}_O$, $\mathbf{L}_t$) with an observational and a theoretical vocabulary. If Beth's Theorem applies, the observable facts stated in $\mathbf{L}_O$ will uniquely fix the denotations of the theoretical terms in $\mathbf{L}_t$, and the latter will be explicitly definable in T. However, this is seldom the case, except for stipulative definitions for mathematical convenience. 

So, let us ask a more general question with no explicit definability assumed. What is the \emph{empirical content} of T, i.e.,  what does it say about the observable world without theoretical superstructure? 

\vspace{0.5ex}

A common syntactic answer has been: the theory T|$\mathbf{L}_O$ of all consequences in the sublanguage $\mathbf{L}_O$ that follow, or are derivable, from T. However, another, and highly influential view, due to \cite{Ramsey1931}, is  semantic: the empirical content of T (its concrete range of application where the theory `works') consists of all $\mathbf{L}_O$-structures that can be \emph{expanded} to a model for T in its full language, or alternatively, the model class MOD(T)|$\mathbf{L}_O$. At least for finitely axiomatized theories T, the empirical content is then defined by the existential second-order \emph{Ramsey sentence} $\exists$ L$_t \cdot$ T($\mathbf{L}_O$, $\mathbf{L}_t$). 

\vspace{1ex}

We always have that the above expandable structures satisfy all principles in the syntactic theory T|$\mathbf{L}_O$. When the model class MOD(T)|$\mathbf{L}_O$ in fact equals MOD(T|$\mathbf{L}_O$), the theory T is called `Ramsey-eliminable'. This is not always the case. 

For a counterexample,  take the first-order theory of the natural numbers $\mathbb{N}$ with $\mathbf{L}_O$ = {<}, and add a unary theoretical predicate A for which it is stipulated that it is non-empty without a lower bound. The $\mathbf{L}_O$-part of this theory is just the first-order theory of < on IN, but the latter standard model cannot be expanded to a model for the whole theory. 

What is always true, however, is that every structure in MOD(T)|$\mathbf{L}_O$ has an \emph{elementary extension} to a model for the whole theory T. However, allowing this move would amount to not just introducing theoretical predicates in an empirical situation, but also additional `theoretical objects'.\footnote{ These notions have also been refined. The empirical content of T might also just be the observational facts and universal laws about observables that follow from T, a universal subtheory of T|$\mathbf{L}_O$.} 

\vspace{1ex}
 
A natural specialization of all this is the case of finitely axiomatized theories T. In that case, the first notion of empirical content should ask for the \emph{strongest consequence} of T in its empirical subvocabulary, which would require \emph{Uniform Interpolation} as defined in \refchapter{chapter:uniform}.\footnote{Interestingly, Craig, p.c., started his seminal work trying to solve the 'strongest consequence problem' for all of first-order logic, and then proving Craig Interpolation as the next best thing.}

\vspace{1ex}

On a more refined view, there can be more layers of vocabulary in a scientific theory: Philosophers sometimes distinguish `observational' (fact-describing), `dispositional' (low-level empirical generalizations) and `theoretical'. But a similar analysis to the one given above applies.\footnote{The distinction also makes sense in natural language, where observable predicates such as hair color or temperature co-exist with more theoretical dispositional predicates such as being ``friendly" or ``breakable", or even highly theoretical predicates such as a person's ``destiny".}

\vspace{1ex}

There is much more to the interface of logic and philosophy of science than the few issues sketched here. For instance, notions of invariance and definability are equally fundamental in both fields, cf. \cite{ButterfieldHaro2025} on the role of symmetries in physics. 

Another crucial aspect of logical analysis that we have left out completely in our brief discussion is the crucial computational role of perspicuous syntax in allowing us to express regularities and laws in scientific theories and compute practical predictions.

\section{Modularity in Knowledge and Belief}		
Moving from scientific theories to   bodies of knowledge and opinion, we now turn to   epistemology as the philosophical study of what it means for humans to know and believe. In this broader field, the crucial role of language returns, if only because what we know or believe is usually expressed in natural language. Various new  issues now emerge, and Craig Interpolation applies in new ways.

\vspace{1ex}

For a start, our knowledge and belief are about certain facts and notions, and the notion of the \emph{topic}  of a proposition, what it is `about', has been a long-standing philosophical theme. 

\vspace{1ex}

Now one natural logical approach to  the topic of a formula defines it as the vocabulary of the smallest sublanguage in which the formula has a logical equivalent. The following result, due to \cite{Parikh2011}, says that this is well-defined.

\begin{theorem} For any formula $\varphi$ stated in a first-order language $\mathbf{L}$, there is a unique smallest sublanguage of $\mathbf{L}$ under inclusion where $\varphi$ has an equivalent.\end{theorem}

Here Craig Interpolation is at work again. Clearly there are inclusion-minimal sublanguages of the finite vocabulary $\mathbf{L}$ where an equivalent can be found. But suppose there were two different ones: $\mathbf{L}_1$, $\mathbf{L}_2$, with equivalents $\varphi_1, \varphi_2$. Then $\varphi_1 \models \varphi_2$, and by interpolation, there is an $\alpha$ in  $\mathbf{L}_1  \cap$ $\mathbf{L}_2$ with $\varphi_1 \models \alpha \models \varphi_2$. But then $\alpha$ would be an equivalent for $\varphi$ in a still smaller sublanguage.

\medskip

Here is another basic language-related issue concerning the modular architecture of first-order theories. It is related to the following equivalent version of the Craig Interpolation Theorem (cf. also the well-known Robinson Joint Consistency Theorem):

\vspace{-0.7ex}

\begin{quote}
\emph{Suppose that a theory} T = T$_1 \cup$ T$_2$ \emph{merging two subtheories is inconsistent, then} \vspace{0.5mm}

\emph{there is a formula} $\alpha$ \emph{in the shared language} $\mathbf{L}_1 \cap \mathbf{L}_2$ \emph{with} T$_1 \models \alpha$ \emph{and} T$_2 \models \neg\alpha$. 
\end{quote}

\vspace{-0.5ex}

Now it is a commonplace that our beliefs are compartmentalized into different topics that can be disjoint. However, most current logics of knowledge and belief just work with one cover-all language without subdivisions. Therefore, it is of interest to \emph{split} the total language into disjoint sublanguages, and ask for the effects of this on the structure of theories. In particular, questions arise about information flow between the various components of a theory T in the total language. 

\vspace{1ex}

Here Craig's Interpolation Theorem has been used in \cite{KourousiasMakinson2007,Parikh2011} to obtain the following result on the fine-structure of bodies of beliefs.\footnote{For convenience, we talk about finitely axiomatized theories though the cited results hold more generally by a simple appeal to the Compactness Theorem for first-order logic.} 

The relevant notion of modularity is the following. We say that a theory T is \emph{split} by a partition $\mathbf{L}_1$, $\mathbf{L}_2$ of its total language $\mathbf{L}$ if there are formulas $\varphi$ in $\mathbf{L}_1$, $\psi$ in $\mathbf{L}_2$ such that T is $\{ \alpha \,| \, \{\varphi, \psi\} \models \alpha \}$. There is a straightforward generalization to larger splittings. 

Next, it is easy to see by two successive applications of Interpolation that the following statement holds whenever 
T is split into $\{ A, B\}$ by a partition $\mathbf{L}_1$, $\mathbf{L}_2$:

\vspace{-0.5ex}

\begin{quote}
If T $\models \varphi$, then there is an $\mathbf{L}_1$-formula  $\alpha$ and an $\mathbf{L}_2$-formula  $\beta$ such that \vspace{0.5mm}

(i) A $\models \alpha$, (ii) B $\models \beta$, and (iii) $\alpha, \beta \models \varphi$.\end{quote}

\vspace{-0.5ex}

A Parallel Interpolation Theorem in this spirit can also be proved for larger finite splittings by suitably iterating Craig Interpolation. Using this, one obtains the following general result, where `largest' refers to the number of partition cells in a splitting:

\begin{theorem} Every theory with a finite vocabulary has a unique largest splitting. \end{theorem}

\vspace{-0.5ex}

As a third example, we just mention a further major theme in formal epistemology, viz.  \emph{information dynamics}. Theories as bodies of knowledge and belief  typically \emph{change} when update takes place with new information produced by observation, reasoning, or communication -- three processes that govern science as well as ordinary life. With the above results in place,  the usual logical accounts of knowledge update and belief revision for rational epistemic agents, \cite{Gardenfors1987,Benthem2011}, can be considerably refined to introduce more topic-sensitivity and modular structure into the representation of what agents know and believe at any stage of the process. 

\medskip

This concludes our survey of uses of Interpolation in various areas of philosophy. Compared to the earlier semantic invariance and dependence theme, we do not have one crisp denominator for all of these, though the pervasive influence of particular vocabularies being used in reasoning and theory construction is an obvious red thread.\footnote{Our focus here has been the \emph{non-logical vocabulary} in formulas. It should be noted, however, that \emph{logical vocabulary} also plays a key role in constructing interpolants, and this feature may go into the intuition that an interpolation theorem requires some sort of expressive completeness for a logical system.}   Whether these two themes unify further than this is not resolved in this paper, and we  leave finding a more integrated perspective as a  desideratum.

\section{A Historical Precursor: Bolzano}
Beth's Theorem and the Interpolation Theorem highlight the importance of language as crucial ingredient in the central notion of logical consequence. But actually, this theme in logic may be less new than we have made it out to be.

A historical precursor to this linguistic  focus is Bernard Bolzano whose ``Wissenschaftslehre'', \cite{Bolzano1837}, has the first systematic semantic notion of consequence, and the first explicit study of different notions of consequence for different purposes (an early form of `logical pluralism'), with varying properties depending on the standards that one adopts for strictness of styles of reasoning. The following discussion is taken from~\cite{Benthem1985}.

\vspace{1ex}

For Bolzano, whether we call a consequence valid depends on which parts of the premises and conclusion we consider as having a \emph{fixed meaning}, and which parts are considered \emph{variable}, and hence amenable to substitutions of other expressions in the same category. 

For instance, a step from premises \{John is taller than Mary, Mary is taller than Paul\} to ``John is taller than Paul'' is valid when we consider ``John, Paul, Mary'' as variable terms, but fix the meaning of ``taller''. But this consequence is not valid if we also consider ``taller'' as a variable term for which we allow  substitutions of other binary predicates that need not be transitive. 

The inference from  ``x is P-er than y, y is P-er than x'' to ``x is P-er than z'' will even remain valid in Bolzano's sophisticated sense if we just fix the meaning of the general comparative construction ``-er'' in natural language which turns adjectives into comparatives.

\vspace{1ex}

Thus, we can view Bolzano's notion of consequence as being \emph{ternary} rather than binary: 

\vspace{-1ex}
\begin{quote} P$_1$, ..., P$_k \models_\mathbf{L}$ C,\end{quote}

\vspace{-1ex}

\noindent  where the vocabulary set $\mathbf{L}$ records which subexpressions of the  formulas involved  we see as variable. 

\medskip

This means that a logical calculus must now keep explicit track of the relevant vocabulary, leading to questions, e.g., about how to combine consequences with different vocabularies. 

For instance, let P$_1$, ..., P$_k \models_{\mathbf{L}_{1}} C$ and Q$_1$, ..., Q$_k \models_{\mathbf{L}_{2}} D$ be given. Should we intersect the vocabulary sets $\mathbf{L}_{1}, \mathbf{L}_2$, or rather take their union when drawing the natural conjunctive conclusion C $\land$ D from the union of the two sets of premises? We leave an answer to the reader.

\vspace{1ex}

All this enriches the usual modern way of thinking about the basic `structural rules' governing styles of inference, where the language is not an explicit parameter: now, it has become crucial.

\vspace{0.5ex}

This explicit manipulation of vocabulary seems in the spirit of the Interpolation Theorem, and ternary sequent calculi are in fact used in proof-theoretic arguments for its truth. But beyond this, we might see ternary sequents that highlight vocabulary as the primary vehicles for inference.\footnote{Not just our theme of language awareness has roots going back to pioneering work in the 19th century. The same is true for our analysis of invariance and definability, which can be traced back as far as the seminal studies in \cite{Helmholtz1883} on the emergence of language as a medium for describing  notions that are invariant under relevant transformations. See also \cite{Weyl1963} for further history. }

\section{Conclusion}
This short and light survey has revolved around  one circle of general ideas, namely, dependence, lifting, vocabulary, transfer, and definability as these occur in various areas of philosophy. What we have tried to show is that both the \emph{notions} and the \emph{results} found in logic show strong analogies with themes in philosophy, and can, under certain circumstances, be applied to yield broader insights.

\vspace{1ex}

It should be noted here that this logic-philosophy interface can be broadened. Our discussion in this chapter has mainly concentrated on the semantic side of logic and model-theoretic analysis. It would also be of interest to revisit all our topics in a proof-theoretic perspective. For instance, the standard proof-theoretic analysis of Craig Interpolation delivers a recursive procedure which, given a proof of an implication $\varphi \rightarrow \psi$ outputs an interpolant and separate proofs for $\varphi \rightarrow \alpha$ and $\alpha \rightarrow \psi$. This changes the discussion from pure  truth or validity to `reasons' for these semantic facts, and this, too, can be connected up with various themes in philosophy.\footnote{In arithmetical terms, switching from validity to provability and coding up formulas and proofs as natural numbers, the complexity of the usual formulation of Craig Interpolation is $\Pi^{0}_2$. The proof-theoretic analysis provides a stronger $\Pi^{0}_1$ version, so there are also interesting technical issues here.}

\vspace{1ex}

But even as it is, our case study also raises issues concerning the fit of logical analysis to philosophy. Logic needs a strictly defined formal \emph{language}, both for its model theory and proof theory, and while this language dependence helps for precision and proving definite results, it may not always fit the spirit of philosophical discourse and thinking, except for those areas which have undergone the twentieth-century `linguistic turn'. And even when language matters, it is not clear whether the restriction to a first-order language (the original habitat of the Beth and Craig theorems)  is always meaningful in philosophy, just as this restriction does not always fit practice  in science. 

In fact, except for some technical subfields in formal epistemology, philosophy of science, and philosophy of language, professional philosophy largely makes do with \emph{natural language}, perhaps enriched with some notations and regimented with fixed technical terminology. What the meta-theorems of logic say in such a setting is clearly a legitimate matter for debate. And we have also suggested occasionally that the properties expressed by these meta-theorems, usually considered a virtue or even a requirement on logic  design, might be less desirable in some areas of philosophy.

\vspace{0.8ex}

Having said this, one can also view the use of logic as offering a model for a philosophical practice (with `model' now in its broader colloquial sense), in which case application just means providing new perspectives and suggestions. Also, the formal similarities noted in this paper could be a force for seeing unifying themes and perspectives across philosophical subdisciplines.

\vspace{0.8ex}

Continuing with the presuppositions of logic, the reliance of Beth's Theorem on a specified \emph{theory} can be questioned, since many philosophical discussions (and perhaps even working practices in sciences such as physics) are about One World where everything of interest takes place.\footnote{Say, in Padoa's method, implicit definability might just be determination in the single model of Euclidean Space that we focus on, the way we think, say, about how a triangle determines its centre of gravity, and to match this, we would need localized versions of Beth's Theorem in suitable structures.} On the other hand, one might also argue that theory reliance is in line with the axiomatic approach in abstract mathematics, and this is definitely the broader context in which Beth saw his own work.

\vspace{0.8ex}

Despite these caveats, we believe that our survey has offered enough points of contact between logic and philosophy along our main themes, and indeed, several of the authors  cited exemplify the union of the two fields, being logicians who were at the same time serious philosophers. More specifically, it might be a rewarding project to  put the relevant literature in the  areas of philosophy mentioned in this survey paper side by side with current logical studies of definability and interpolation in a much more detailed and systematic manner. And our selection of topics provided just one sample of course, while even the topics selected  could  profit from broadening the logic interface to include themes from computational logic such as complexity, from interfaces with quantitative approaches, \cite{BenthemIcard2023}, or from the vast area of non-classical logics. 

\vspace{1ex}

Finally, as to what is at stake,  the real benefit of the juxtaposition in this paper   may not be so much logical solutions to philosophical problems as \emph{mutual inspiration} – since logic can also learn from the philosophical imagination and wide-ranging theorizing in the areas  highlighted here.

\medskip

\section*{Acknowledgments} I thank Alexandru Baltag, Balder ten Cate, S\"{o}ren Brinck Knudstorp,   Martin Otto, Ian Pratt-Hartmann, Pablo Rivas Robledo, Martin Stokhof, Albert Visser, Frank Wolter, and Urszula Wybraniec-Skardowska, as well as audiences for presentations of this material in Amsterdam, Beijing, Berkeley, and Dortmund for their comments on  topics in this paper. 

\addcontentsline{toc}{section}{Acknowledgments}


\bibliography{BIB}
\end{document}